# Performance Analysis of Dynamic Source Routing Protocol

[1] Amer O. Abu Salem, [2] Ghassan Samara, [3] Tareq Alhmiedat

[1] College of Science & Technology Zarqa University, Zarqa-Jordan
[2] College of Science & Technology Zarqa University, Zarqa-Jordan
[3] College of IT Tabuk University, Tabuk, KSA

## ABSTRACT

Dynamic Source Routing (DSR) is an efficient on-demand routing protocol for mobile ad-hoc networks (MANET). It depends on two main procedures: Route Discovery and Route Maintenance. Route discovery is the procedure used at the source of the packets to discover a route to the destination. Route Maintenance is the procedure that discovers link failures and repairs them. Route caching is the sub procedure serviceable to avoid the demand for discovering a route or to reduce route discovery delay before every data packet is sent. The goal of this paper is to evaluate the performance of DSR. Different performance expressions are investigated including, delivery ratio, end to-end delay, and throughput, depending on different cache sizes and different speeds. All of that as a study to develop a new caching strategy as a future work.

## 1. INTRODUCTION

A Mobile Ad-hoc Network (MANET) is the active research topic in wireless communications. It is a collection of two or more nodes in which the communication links are wireless; without the service of any fixed infrastructure or centralized administrator. The network is Ad-hoc because each node is able to receive and forward data to other nodes, and so the judgment of which nodes forward data is made dynamically based on the network connectivity. The advantage of this type of network is that it does not require any kind of infrastructure, like a base station in cellular network, so the most important challenge in a mobile Ad-hoc network is to implement routing protocols that can manage these network topology changes to maintain and rebuild the dependable routes in a timely approach.

This paper has been structured as follows. Section 2 gives an overview of routing technique. Section 3 describes the cache scheme used and the problem statement. In section 4 we summarize the route caching in DSR protocol. Section 5 describes the scenario simulation which used to evaluate the performance of DSR. Section 6 summarizes the results of the NS-2 simulator and the evaluation of that case scenario, and finally, in section 7 we demonstrate our conclusions.

## 2. ROUTING

Several various routing algorithms for Ad-hoc networks, with their particular advantages and disadvantages have been proposed until now. Researchers traditionally classify these protocols as proactive protocols, reactive protocols, or hybrid of them, based on the algorithms that find new routes or update existing ones. Proactive routing is implemented by exchanging routing tables, (such as SDV, WRP). Reactive routing is on demand routing, (such as DSR, AODV). It has been shown that reactive routing is more suited for Ad-hoc than the proactive one. In reactive routing, there is two main phases: routing discovery and routing maintenance. The routing discovery phase is based on request and reply recycles, so the cost is high, which will decrease the performance of network. The investigators use three primary strategies to decrease discovery cost: - Enhancing cache, every node has a local cache to save the rout path from itself to destination (such as DSR, AODV). - Local flooding, use broadcast flooding to specific neighbours based on specific regulations. The flooding broadcast locally will reduce the discovery cost (such as LAR, ZRP).– Multiple route path, using multiple paths to transmit data parallel or concurrently, option path will reduce the number of requests (such as SMR, AOMDV). On-demand routing protocols achieve better than table driven routing protocols in mobile Ad-hoc networks. In an on-demand strategy, node attempts to discover a dependable route when it wants send data packets to a destination. To avoid the cost of discovering a route for each data packet, nodes store that discovered routes in a cache. Thus, an effective caching strategy is an essential part of any on-demand routing protocol for wireless Ad-hoc networks to keep a cache up to date always.

## 3. CACHING

A cache scheme in reactive routing protocols is defined by the following set of design issues that determine cache management strategy; the 1st issue is a store policy; which it is the set of rules that determine the cache information structure, to be suited to the route cache on demand. The researchers proposed and studied two different cache structures, called path cache and link cache, and applied them to DSR [5]. A path cache is structured that each cache entry acts a full path from a source to all destinations, while a link cache has each individual link to one destination. So, a link cache has the possible to use the route information more efficiently. The 2nd issue is a read policy; the rules that determine when use a cache entry, and decide what is the reaction when sending a new message from the source node. For example, DSR protocol depends on several policies: cache replay, to reply message with information about a route request that stored in cache of intermediate nodes, the second policy is a salvaging reply, when data packet meet a broken link in the route path, the intermediate node can use alternative path to destination from its own cache, finally, volunteer reply, a node can listens for packets not directed to it (packet sniffing), if the node has a better route to the destination node of a packet, it sends a message reply to the source node with this new better route. The third issue is a writing policy; the rules for determining when to write an entry into the cache, and which information have to be cached. The main trouble for the writing policy is guarantee to cache valid paths. The last one is a deletion policy; it is the rules for determining when to delete an entry from the cache and which information has to be deleted from the cache. Deletion policy is really the most critical part of a cache scheme. Two kinds of "errors" can happen: Early deletion, a cached route is deleted when it is still valid. And late deletion, a cached route is not deleted even if it is no longer valid. As DSR cache is based on path organization strategy, which depend on receive route reply packets by intermediate or source node to cache this path without any processing, and it will delete whole path when a data packet meets a broken link by receiving error packets, and there is no automatic link expiration strategy. These will result Inefficient Cache Organization.

## 4. ROUTE CACHING IN DSR

DSR- Dynamic Source Routing protocol is an on-demand routing protocol, which means that nodes request routing selective information only when demanded. DSR is depending on source routing concept, where the sender creates a source route in the header of the packet. This source route includes all the addresses of the intermediate nodes responsible of forwarding the packet to the destination. When a sender decides to communicate with another destination, it checks into its route cache to discover if there is any routing information related to that destination. If route cache does not include any such information to that destination, then the sender will initiate a route discovery process by broadcasting a route request. If the route discovery is successful, the sender receives a route reply packet contains a sequence of network intermediate nodes through which it may reach the destinations. Nodes may reply to requests even if they are not the destination to reduce the delay time. It is also possible that intermediate nodes which pass on the packets can overhear the routes by examining the packet and learning about routes to confident destinations. DSR uses path caches or link caches. In a path cache, a node stores each route starting from itself to another node. In a link cache, a node adds a link to a topology graph, which represents the node's view of the network topology. Links obtained from different routes can form new routes. Thus, link caches provide more routing information than path caches. DSR also implements a route maintenance process, which uses the data link layer references to learn of any lost links. If any lost link was detected, a route error control packet is sent to the initiating node. Consequently, the node will remove that hop in error from the node's route cache, and all routes that contain this hop must be truncated. To improve the DSR performance we studied the primary and the secondary cache used in DSR with path caches, without any modification to the protocol algorithm. DSR uses a secondary cache. We use the primary cache to holds active routes that were requested by that node to send data. The secondary cache to store the routes a node overhears. And If a route in the secondary cache is used to respond to a request route or to send packets, it will be added to a primary cache. And use a secondary cache to store both the routes a node overhears and the overheard routes learned from reply routes.

## 5. SIMULATION SETUP

This research was conducted using NS-2. NS-2 is chosen as the simulation tool among the others simulation tools because NS-2 supports networking research and education. Ns-2 is suitable for designing new protocols, comparing different protocols and traffic evaluations. The simulated network consisted of 50 nodes randomly spread in a 300x600m area at the start of the simulation. The tool setdest [10] was used to generate mobility scenarios, where nodes are moving at five different uniform speeds with range between 0 to 15 m/s with a margin of ±1 and a uniform pause time of 10s. We simulated the steady-state conditions of the network with CBR traffic model; these were generated using the tool cbrgen.tcl [10], with the following parameters: CBR: Constant Bit Rate traffic model. This was generated at a deterministic rate with some randomizing on the inter packet departure interval. Packets size was set to 64 bytes generated at a constant rate of 2 kb/s.

Next table summarizes the simulated network area topology, mobility parameters, and the data traffic scenario used in the simulation.

**Table 1:** Area Topology, node's mobility and traffic model

| Parameters | Value |
|---|---|
| Topology Area | 300*600 m |
| Number of Nodes | 50 |
| Node Transmission range | 100m |
| Total Simulation Time | 1000 s |
| Bandwidth | 2 Mb/s |
| Pause Time (Uniform) | 10 seconds |
| Speed (Uniform) | 0,1,5,10,15 ± 1 m/s |
| Traffic model | CBR |
| Packet size | 64 byte |
| Rate | 2 kb/s |

## 6. RESULTS AND DISCUSSION

In DSR, the cache implementation of the nodes is classified into two parts: primary and secondary. Generally, DSR node uses the primary cache to holds active routes that were requested by that node to send data. The secondary cache holds routes that were overheard from other nodes. Using the scenario and communication models which determined above, we changed the size of the primary and secondary caches defined in dsr/mobicache.cc [10] and changed the speed of the nodes, to observe its effect on several metrics in analysing the performance of routing protocol. These metrics are as follows:

### 6.1 Data Packet Delivery Ratio
Total number of delivered data packets divided by total number of data packets transmitted by all nodes [1]. This performance metric provide us an estimate of how well the protocol is doing in conditions of packet delivery at different speeds using CBR traffic model and different cache sizes.

### 6.2 Average End-to-End Delay (seconds)
The average time a data packet takes to access the destination. This metric is calculated as: The time at which first data packet arrived to destination – The time at which first packet was transmitted by source. This includes all possible delays caused by buffering for the duration of route discovery latency, queuing at the interface queue, retransmission delays at the MAC, propagation and transfer times. This metric is necessary to understand the delay which introduced by path discovery.

### 6.3 Average Throughput (messages/second)
Average Throughput (messages/second) is the average rate of successful packet delivery over a communication channel, this metric is calculated as:

The average total number of delivered data packets divided by the total duration of simulation time. We study the throughput of the protocol in terms of average of number of messages delivered per one second. The simulation traces were analysed, the following are the observations mentioned:

### 6.4 Data Packet Delivery Ratio
Figure 1 shows Data packet delivery ratio versus cache size for the different speeds. The results indicate that the packet delivery ratio decreases as the cache size decreases below 5 for primary and 10 for secondary, for all speeds. The plot indicates that there is not much change in the packet delivery ratio for cache sizes of 10 or above especially when speed is more than 10m/s. This experiment shows that if we want to reduce the cache size in the nodes, we should probably not reduce it below 5 for Primary and 10 for secondary for all speeds and not over 10 for primary and 20 for secondary for high node speed which is equal to 15m/s and more, as that would affect the performance of the network.

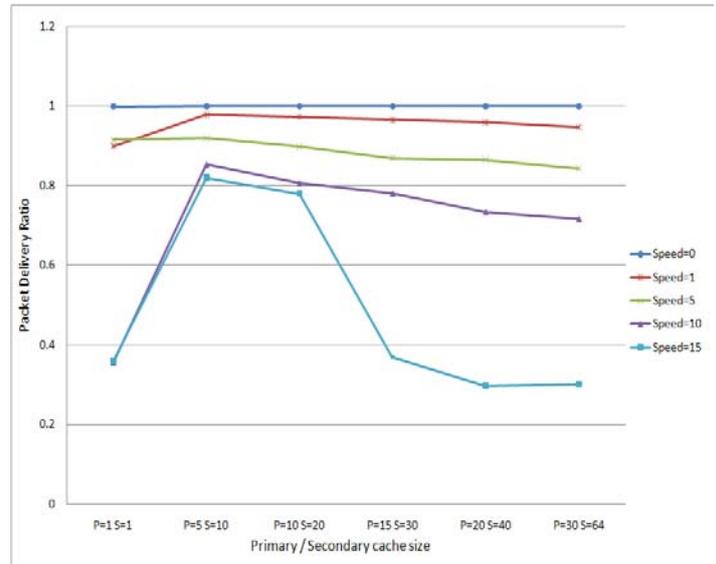

**Fig 1:** Data packet delivery ratio vs. cache size

### 6.5 Average End-to-End Delay
Figure 2 illustrates end-to-end delay versus cache size for the different speeds. DSR has maintained a low delay as well for speeds less than 10 m/s for any cache size. However, a dramatic increase in delay was observed at higher speeds, the reason is that at high mobility. It was observed that at higher speeds which equal to 15 and more introduces a relatively highest delay except when primary cache size and secondary cache size between (5, 10) and (10, 20) for CBR traffic models.

### 6.6 Average Throughput (Messages/Second)
Figure 3 shows the average of throughput of the protocol measured in messages/second versus cache size for different speeds. DSR has maintained a high throughput at speeds less than 5 m/s. This was due to the use of route cache and overhearing properties of DSR routing protocol. On the other hand, the throughput observed when speed greater than or equal 10m/s using the CBR traffic model was low , especially when primary cache size greater than 10 for the speed equal 15m/s.

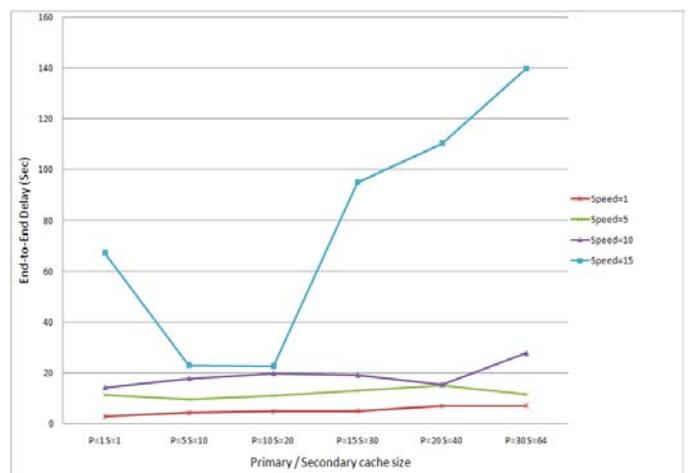

**Fig 2:** Average end-to-end delay vs. cache size

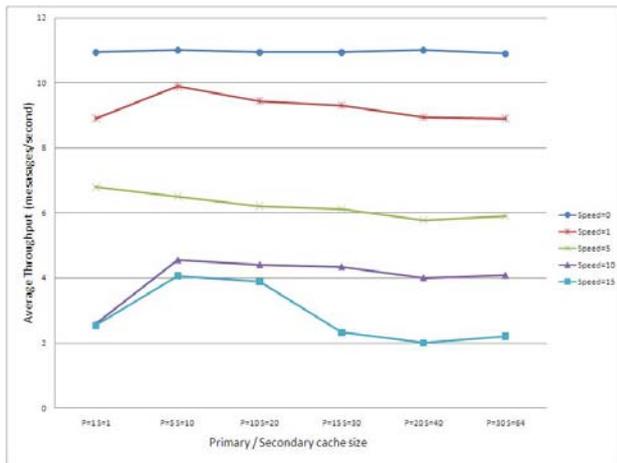

**Fig 3:** Average throughput vs. cache size

## 7. CONCLUSIONS

This paper focuses on the analysis and performance of DSR protocol, depending on studying two factors; the cache size and the speed by simulation using NS-2 with nodes have cache divided into two parts (primary and secondary) changing at sizes ranging from (p=1, s=1) to (p=30, s=64) and moving at speeds ranging from 0 to 20 m/s. The DSR routing protocol has acceptable performance in terms of data packet delivery ratio, throughput and This paper focuses on the analysis and performance of DSR protocol, depending on studying two factors; the cache size and the speed by simulation using NS-2 with nodes have cache divided into two parts (primary and secondary) changing at sizes ranging from (p=1, s=1) to (p=30, s=64) and moving at speeds ranging from 0 to 20 m/s. The DSR routing protocol has acceptable performance in terms of data packet delivery ratio, throughput and end-to-end delay at speeds less than 15m/s. Furthermore, the cache size (both the primary and secondary) has not played a crucial role in the networking performance. Whereas at speeds more than 15m/s, the cache size expressed a remarkable role; the greater the cache size the greater the end-to-end delay will be, and vice-versa. Based on this result, the best cache size in case of high speeds should not be more than 10 for primary cache and 20 for secondary cache. As a continuation of this research work, it would be very interesting to develop a new caching strategy, and evaluate other protocols that have important operations in MANET.

## REFERENCES

[1] H. AhleHagh, and W. R. Michalson, "Statistical Characteristics of Wireless Network Traffic and Its Impact on Ad Hoc Network Performance," in Advanced simulation Technologies Conference, Orlando, USA, pp. 66-71, April 2003

[2] Azizol Abdullah, NorlidaRamly, Abdullah Muhammed, Mohd Noor Derahman: Performance Comparison Study of Routing Protocols for Mobile Grid Environment, pp 82-88, IJCSNS International Journal of Computer Science and Network Security, VOL.8 No.2, February 2008

[3] A. Al-Maashri, M. Ould-Khaoua, "Performance Analysis of MANET Routing Protocols in the Presence of Self-Similar Traffic" Proceedings of the 31st IEEE Conference on Local Computer Networks, Florida, USA, November 2006

[4] Y. Chen, X. Guo, "An Adaptive Routing Strategy Based on Dynamic cache in Mobile Ad Hoc Networks", ISPA 2004, LNCS 3385, pages 357-366, 2004

[5] Zhao Cheng, Wendi B. Heinzelman: Adaptive local searching and caching strategies for on-demand routing protocols in ad hoc networks.IJHPCN 4(1/2): 51-65, 2006

[6] D. Johnson, D. Maltz, Y Hu and J Jetcheva. The Dynamic Source Routing Protocol for Mobile Ad Hoc Networks. IETF MANET Internet Draft, Feb 2002.

[7] M. Frodigh, P. Johansson, P. Larsson, "Wireless Ad-hoc networking: The art of networking without a network" Ericsson Review, No. 4, pages 248-263, 2000.

[8] W. Lou, Y. Fang, "Predictive Caching Strategy for On-Demand Routing Protocols in Wireless Ad Hoc Networks", Wireless Networks volume 8, pages 671-679, 2002

[9] The Network Simulator NS-2 tutorial homepage, http://www.isi.edu/nsnam/ns/tutorial/index.html

[10] MANET Simulation and Implementation at the University of Murcia, Available from http://masimum.dif.um.es/, 2006